\documentclass[sigconf,nonacm]{acmart}
\usepackage{tikz}
\AtBeginDocument{%
  \providecommand\BibTeX{{%
    \normalfont B\kern-0.5em{\scshape i\kern-0.25em b}\kern-0.8em\TeX}}}

\makeatletter
\def\@ACM@checkaffil{% Only warnings
    \if@ACM@instpresent\else
    \ClassWarningNoLine{\@classname}{No institution present for an affiliation}%
    \fi
    \if@ACM@citypresent\else
    \ClassWarningNoLine{\@classname}{No city present for an affiliation}%
    \fi
    \if@ACM@countrypresent\else
        \ClassWarningNoLine{\@classname}{No country present for an affiliation}%
    \fi
}
\makeatother

\begin{document}

\title{S3C2 Summit 2025-03: \\ Industry Secure Supply Chain Summit}

\author{Elizabeth Lin$^{*}$, Jonah Ghebremichael$^{*}$, William Enck$^{*}$, Yasemin Acar$^{\dagger}$, Michel Cukier$^{\ddagger}$, \\ Alexandros Kapravelos$^{*}$, Christian Kästner$^{\mathsection}$, Laurie Williams$^{*}$}

%% authors that will appear in the ACM reference format (without authormarks)
\def \authors{Elizabeth Lin, Jonah Ghebremichael, William Enck, Yasemin Acar, Michel Cukier, Alexandros Kapravelos, Christian Kästner, Laurie Williams}

\affiliation{%
    \institution{ $^*$North Carolina State University, Raleigh, NC, USA}
}
\affiliation{%
    \institution{$^\dagger$Paderborn University, Paderborn, Germany and George Washington University, DC, USA}
}
\affiliation{%
    \institution{$^\ddagger$University of Maryland, College Park, MD, USA}
}
\affiliation{%
    \institution{ $^\mathsection$Carnegie Mellon University, Pittsburgh, PA, USA}
}
%\email{{firstname.surname}@uni-paderborn.de}
%\email{mcukier@umd.edu}
%\email{{whenck, akaprav, lawilli3}@ncsu.edu}
%\email{kaestner@cs.cmu.edu}

\renewcommand{\shortauthors}{Secure Software Supply Chain Center (S3C2)}
\renewcommand{\shorttitle}{S3C2 Summit 2023-06: Government Secure Supply Chain Summit}

\begin{abstract}
  
  Software supply chains, while providing immense economic and software development value, are only as strong as their weakest link. Over the past several years, there has been an exponential increase in cyberattacks specifically targeting vulnerable links in critical software supply chains. These attacks disrupt the day-to-day functioning and threaten the security of nearly everyone on the internet, from billion-dollar companies and government agencies to hobbyist open-source developers. 
  The ever-evolving threat of software supply chain attacks has garnered interest from both the software industry and US government in improving software supply chain security. 
  
  On Thursday, March 6th, 2025, four researchers from the NSF-backed Secure Software Supply Chain Center (S3C2) conducted a Secure Software Supply Chain Summit with a diverse set of 18 practitioners from 17 organizations. 
  The goals of the Summit were:
  (1)~to enable sharing between participants from different industries regarding practical experiences and challenges with software supply chain security;
  (2)~to help form new collaborations;
  and 
  (3)~to learn about the challenges facing participants to inform our future research directions. 
  The summit consisted of discussions of six topics relevant to the government agencies represented, including software bill of materials (SBOMs); compliance; malicious commits; build infrastructure; culture; and large language models (LLMs) and security. 
  For each topic of discussion, we presented a list of questions to participants to spark conversation. 
  In this report, we provide a summary of the summit. 
  The open questions and challenges that remained after each topic are listed at the end of each topic's section, and the initial discussion questions for each topic are provided in the appendix. 
  
\end{abstract}

\iffalse

\fi



%\received{30 September 2022}
%\received[revised]{1 December 2022}

\maketitle

\begin{tikzpicture}[overlay, remember picture]
\node[anchor=north west, %anchor is upper left corner of the graphic
      xshift=17.5cm, %shifting around
      yshift=-2.1cm] 
     at (current page.north west) %left upper corner of the page
     {\includegraphics[width=2.1cm]{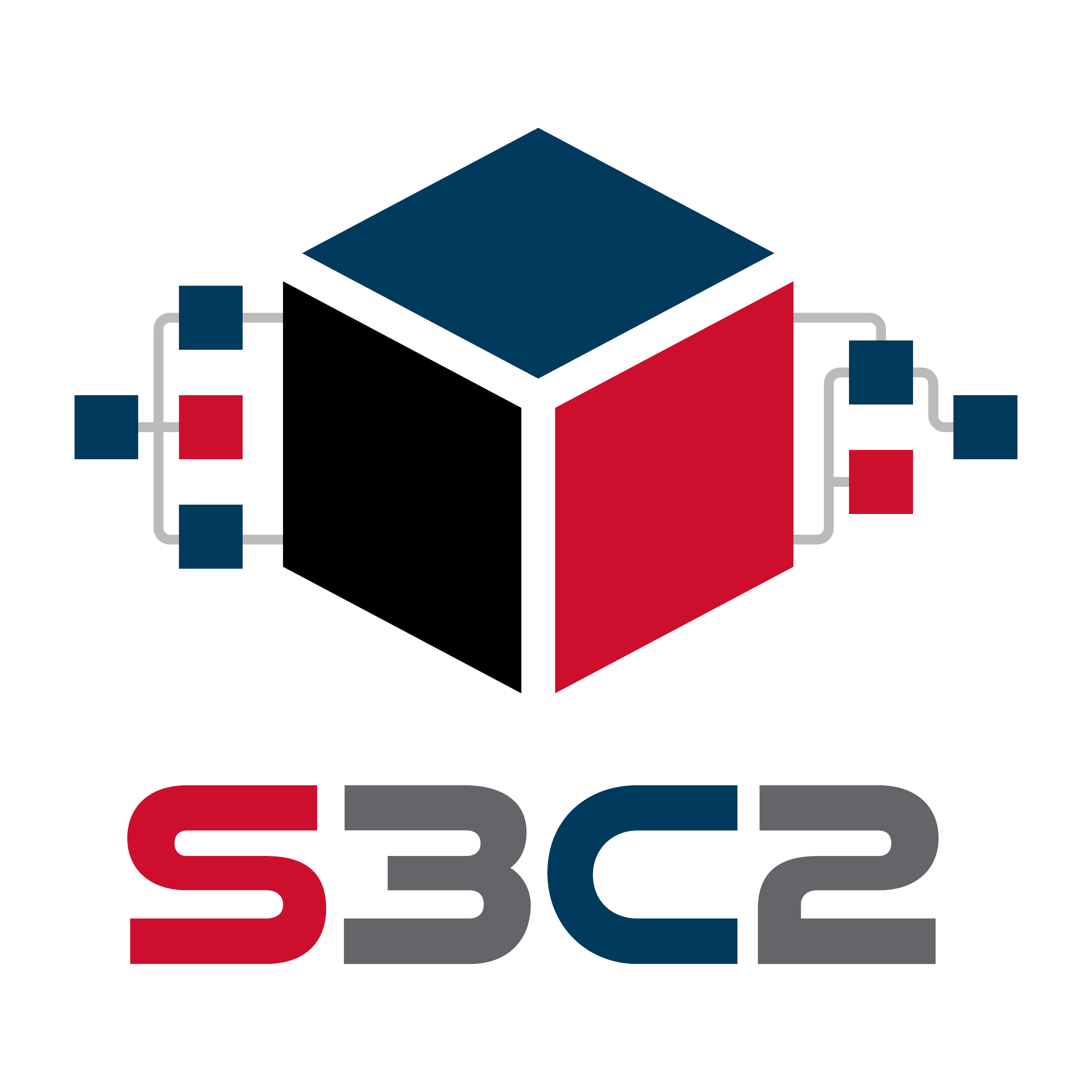}}; 
\end{tikzpicture}

\section{Introduction}
\label{sec:intro}
On Thursday, March 6th, 2025, four researchers from the NSF-backed Secure Software Supply Chain Center (S3C2)\footnote{https://s3c2.org/} conducted a day-long Secure Software Supply Chain Summit with a diverse set of 18 participants from 17 companies. 
The goals of the Summit were to:
  (1)~to enable sharing between participants from different industries regarding practical experiences and challenges with software supply chain security;
  (2)~to help form new collaborations;
  and 
  (3)~to learn about the challenges facing participants to inform our future research directions. 

The Summit was conducted under the Chatham House Rule, which means that all participants can freely use the information discussed. However, disclosing who was present, their affiliations, or who said what is forbidden. 
As such, this report also adheres to the Chatham House Rule, and the affiliations of the participants are not disclosed.  
Summit participants were recruited from 17 companies invested in software security. 
Attendance was intentionally capped to create an environment that encourages candid conversations among key stakeholders. 

The Summit consisted of discussions of six topics that were decided ahead of time by the participants through voting on which topics to discuss.
The voting process ensured that the topics were of interest and relevant to the companies represented. 
The discussion topics included software bill of materials (SBOMs), compliance, malicious commits, build infrastructure, culture, and large language models. 
Each topic was moderated by one of the S3C2 researchers, beginning with a list of questions to spark conversation.  These questions are provided in the appendix. For each of the six topics, this report provides a discussion of continued themes from prior summits \cite{Summit1, Summit2, Summit3, Summit4, Summit5, Summit6, Summit7}, followed by new ideas that emerged in this particular summit.  

Four S3C2 researchers (two professors and two PhD students) took notes on the discussion. The PhD students created a first draft of this report based on these notes, which the two professors then reviewed and revised. 
The draft was then reviewed by the Summit participants and the other authors of this report, who are also S3C2 researchers and experts in software supply chain security. 

The remaining sections of this report summarize the Secure Software Supply Chain Summit.

\section{Software Bill of Materials}
\label{sec:sbom}

The first topic of discussion was Software Bill of Materials.
A Software Bill of Materials (SBOM) is a nested inventory of “ingredients” that make up the software component or product and helps to identify and keep track of third-party components of a software system.
The Executive Order (EO) 14028 states that any company that sells software to the federal government must issue a complete SBOM that complies with the National Telecommunications and Information Administration (NTIA) Minimal Elements~\cite{EO}.

\subsection{Continued Themes}
\label{ssec:sbom-checkbox}

\paragraph{SBOM as a Checkbox}

One of the key takeaways from our initial summits in 2021~\cite{top-five-spmag22} was the desire to utilize SBOMs to enhance security, rather than simply serving as a checkbox.
Four years later, this fear was still a topic of discussion.
Participants pointed out that some companies do not want any questions on their produced SBOMs and would sometimes ``greenwash'' the SBOM results, making it seem as if there are no security issues within their software.
This sentiment was intertwined with a significant concern about the \emph{trust} and \emph{accuracy} of SBOM results.
There was a growing concern that developers checking off the SBOM box without giving much thought to its accuracy would undermine its value.

Similar to prior summits, some attendees argued that the real importance of SBOMs lies in the \emph{process}: generating and consuming SBOMs provides an opportunity to better understand software components.
They argued that companies willing to provide SBOMs are likely to be more transparent with their software and more willing to address any security issues that arise.
SBOMs as a checkbox create an opportunity for companies to communicate and discuss any security issues. During this process, you can assess the company's stance and attitude toward security in its products.

\paragraph{Trusting SBOMs}

Participants noted that SBOMs received from a supplier may differ from what was expected in the software.
SBOMs from different tools may also return different results.
Companies face challenges in resolving these differences and are unsure how much trust to place in SBOM results.
An attendee expressed the opinion that the software industry should decide on \emph{a} tool so there is no need to resolve differences.

\paragraph{SBOM Sharing}

Participants shared that they are still unsure of how to generate and consume SBOMs.
There is high-level guidance from the EO and corporate, but it is not clear to developers how to implement that.
Communicating SBOM information across software suppliers and consumers is challenging.
A participant shared that they have yet to receive many requests from customers for SBOMS, but they have been using SBOMs internally to help keep track of components in the company.

\paragraph{Legacy Software}

%Sharing SBOMs requires a company to inventory its software component, and that itself is a challenge.

A participant from a large organization shared their challenges with generating SBOMs for legacy software.
Large organizations frequently maintain dozens of products and applications. 
Many of these products are from companies they acquired, or companies that those companies acquired.
These products are also spread across many different business units.
Each product has its own supply chain and individual SBOM.
To further complicate SBOM creation, products are often more than a dozen years old and code from embedded systems, making it hard to produce a bill of materials.

\subsection{New Ideas}

\paragraph{Addressing SBOM Noise}

Participants discussed the challenge of noise in SBOMs.
Early forms of SBOMs only keep track of static components. 
However, SBOMs have expanded to include dynamic components during the build process.
These processes can generate SBOM noise.
Participants expressed a need for \emph{contextual SBOMs} (generated by monitoring the build process) and \emph{core SBOMs} (information on the core software components).
With build processes, the output can be different when run in different environments.
A participant expressed a desire to bring back hermetic builds (building with no network access) into SLSA \cite{SLSA} so there is more confidence in the build process.
For example, an SBOM declaration can be used to prefetch dependencies, thereby ensuring the SBOM is accurate.
However, some participants argued that hermetic builds are a challenge in themselves.

%The accuracy of SBOM tools was another challenge mentioned by attendees.
%Different package managers have different ways of including and resolving libraries.
%Attendees reported certain tools sometimes better at resolving components in certain ecosystems.

\section{Compliance}
\label{sec:compliance}

The second topic of discussion was compliance.
Compliance standards such as SSFD \cite{SSDF} define secure software development practices.
The EO requires companies supplying software to the federal government to attest that the software is securely developed by completing a self-attestation form.

\subsection{Continued Themes}

\paragraph{Legacy Software}

As SBOMs are often central to compliance requirements, themes from the SBOM panel continued into the compliance panel.
One panelist shared their experiences attempting to self-attest to a build pipeline that has been working for over 30 years.
Gathering the necessary documentation to meet the attestation was a nightmare.
They also encountered tooling problems.
For example, they have a Java project that uses Ivy, and the tools they wanted to use supported Maven, not Ivy.
They eventually found a tool that supported Ivy, but it did not work because they had customized Ivy too much for the project.
These types of problems are common for older projects.
As noted in the SBOM panel, mergers and acquisitions result in collections of products that have a wide range of nuances, all of which need to be addressed.

\paragraph{Value of Compliance}

There was a lengthy discussion about the extent to which self-attestation has improved security compared to a checklist approach.
In the current self-attestation approach, the company CEO is required to attest that a secure software process was used, which raises security to the top levels of the executive chain.
Some participants concluded that the combination of getting a company to change, along with a culture change, is the benefit of self-attestation over a checkbox, and it will win every time.
However, other participants reflected that the current self-attestation declares compliance "to the best of [the CEO's] knowledge," and therefore leaves room for interpretation.
Participants also noted that alternative models such as ISO and FedRAMP are more focused on getting better versus passing.
For example, FedRAMP at the end of the day is always a negotiated finding.

Finally, there was a sentiment that self-attestation was healthy for the company, but not healthy for the teams enforcing it.
It creates more work for the teams as attestation is a strict yes/no response.
They argue it would be healthier for the teams if attestation could be expanded to acknowledge the standards and commit to improving development practices if attestation is not met.

\paragraph{Minimum Vialable Compliance}

There was a sentiment that many companies are doing just enough to meet compliance and no more.
They are also identifying subtle loopholes.
For example, one participant shared the trick of not bumping the version of a product, because if a product's release was before a certain date, it does not require an attestation.
There was a sentiment that this loophole was commonly known.

\subsection{New Ideas}

\paragraph{Changes in SBOM Sensitivity}

In prior summits, many participants raised concerns about making SBOMs public, citing concerns over intellectual property and giving attackers a better roadmap to attack their systems.
This sentiment is beginning to change as SBOMs are becoming more common.
One participant noted that we should not be fearful of adversaries getting information from SBOMs, because the tools to get this information from binaries are getting increasingly faster and better.
If this sentiment grows, there may be more open sharing of SBOMs.

\paragraph{Changes Cause Extra Work}

A participant shared that they had initial legal requirements that were written for Java code.
However, when they transitioned to Golang, the legal requirements did not apply.
Changing compliance standards is also a headache for developers.
A participant shared that they initially followed the SLSA v0.1 standard and put in effort to comply with it.
Then, when SLSA updated the specifications to v1.0, they had to change the way they implemented the standards, which was a challenge.
Complying with the standards was also a significant hurdle.
Thus, a participant shared that they developed automated pipelines to reduce work for developers.

\paragraph{Differences Across Companies and Industries}

When participants shared development practices within their company, we observed that different companies practice compliance in different ways.
One participant shared that at a previous company, code was proprietary and highly-restricted, but that is not the case at the current company.
Different companies have different standards for \emph{code} within the company.
Some treat application code as restricted data and are treated with the highest standard of security, while others do not.

The standards for compliance vary across different industries.
Most participants were from the tech sector and were familiar with standards in the tech industry.
One participant was from the energy sector and provided insight into different standards.
The Department of Energy has a program, CyTRICS,\footnote{https://cytrics.inl.gov/} specifically for attestation in the energy sector.

\section{Malicious Commits}
\label{sec:malicious-commits}

Malicious commits were the next topic of discussion.
The XZ utils backdoor (discovered in early 2024) was the result of malicious commits.
The malicious actor gained the trust of the repository owner by starting with benign commits to the XZ utils library.
After gaining trust, the malicious actor introduced the backdoor through a malicious commit.
Since the XZ utils backdoor, social engineering and malicious commits have been a frequent topic of discussion.

\subsection{Continued Themes}

\paragraph{Intent}
Participants discussed approaches to identifying malicious commits.
Determining whether a commit has malicious intent is a challenge.
Intent can only be inferred from the outcomes of the commit.
Participants noted that a commit may not always have malicious intent, but it could still include bugs or vulnerabilities.
Software bugs and malicious intent can both cause issues in software; therefore, it is challenging to differentiate between them.

\paragraph{Zero-Trust}
The topic of trust was brought up when discussing the intent of commits.
One participant shared that they have a \emph{zero-trust} policy for all commits, they assume that there could be malicious intent in all commits, other participants echoed the same opinion.

The topic of trust in your upstream was also discussed.
Upstream software is also vulnerable to malicious commits, thus participants discussed that there should also be zero-trust in upstream software.
Finding an anchor point for trust is a challenge as well, participants discussed the concept of trusting trust from Ken Thompson.
The discussion also touched on having a web of economics that benefits and supports upstream maintainers to prevent incidents like the XZ utils backdoor from happening.

\paragraph{Countermeasures}
Various countermeasures were mentioned when discussing malicious commits.
As all code can be vulnerable to malicious commits, a participant expressed the view that sandboxing is a defense that should be implemented.
Another participant proposed a penalty for malicious commits to deter actors with malicious intent.
As the XZ utils backdoor was introduced due to the library only having a main maintainer, the malicious actor was able to take advantage of it and easily gain the trust of the maintainer.
Another countermeasure mentioned by a participant is to implement multi-party checks for commits, removing the risk of a single point of failure.

\subsection{New Ideas}

\paragraph{Commits are a point in time}
When discussing malicious commits, a temporal aspect was introduced in the discussion.
A participant shared their view that a commit is a single point in time, it is difficult to determine the intent of a single commit.
Looking at each commit in isolation does not reveal much. 
However, when you look at all commits from a user over time, then there is more information to determine the intent of the user.

\paragraph{Addressing Build Scripts}
The attack on the XZ utils library was also reflected in the discussion.
In this attack, the build scripts were maliciously modified to inject a backdoor into the compiled artifact.
One participant noted,  ``build scripts is where everyone has blind spots.''
Call graph analysis will not work on build scripts.
Builds are also monolithic environments where fine-grained sandboxing within the build is rarely considered.

\section{Build Infrastructure}
\label{sec:build-infra}

Various build platforms and CI/CD tools support developers in automating software development processes, including building, testing, and deployment. 
Build infrastructure enhances the integrity of software builds by creating documented and consistent build environments, isolating build processes, and generating verifiable provenance. 
Additionally, reproducible builds contribute to this integrity by making the build system entirely deterministic and verifiable, ensuring builds can be consistently reproduced and verified.

\subsection{Continued Themes}

\paragraph{Reproducible Builds}
The topic of reproducible builds was brought up early in the discussion.
Some participants believe reproducible builds are possible if the work is put in.
With reproducible builds, most requirements of the SLSA model can be achieved.
However, there are some challenges with it.
Inventory was one of the first points discussed.
Knowing what is in the pipeline is important for reproducible builds.
Having a controlled environment was another discussion point.
Different build environments could generate different build outputs.
If all builds have the same controlled environment, this would reduce the variances in the outputs.
Even with the same build environment, there are still other challenges.
An example brought up by participants is timestamps; timestamps inserted during the build process create different build outputs.

Other participants struggle to see the value in reproducible builds.
They shared that having the same output from two different builds is interesting, but they don't see the value in it.
They believe that build provenance, verifiable data about a build, is valuable.
One participant also pointed out that if reproducibility is a challenge, we should at least aim for rebuildability.

The discussion then circled back to a discussion point we mentioned in Section~\ref{ssec:sbom-checkbox}.
Similar to discussing whether there is value in SBOMs, participants discussed the benefit of investing in reproducible builds.
While it is a challenge to make builds reproducible, attention and discussion on this topic will motivate more companies to improve their build processes.

\subsection{New Ideas}

\paragraph{Observing Builds}
Another topic of discussion was the observability of builds.
By observing builds, developers can identify anomalies in the build process that could lead to tampering.
Participants discussed how to \emph{observe} the build process.
Some participants pointed out that referring to historical data to detect potential anomalies, and having an automated process that automatically detects differences from historical data.
Participants also discussed that while having an automated script do the work is nice, observing and monitoring the builds still requires human effort.
A developer has experience and knowledge to triage whether a change is a potential anomaly, something an automated script may not capture.
To observe builds, participants discussed keeping all build data available as long as a project is alive, which involves a large amount of data.

\paragraph{Managing AI Models}
Other topics were also brought up during the build infrastructure panel.
A participant mentioned AI models, which could also be using other AI models, can be hard to manage.
Using a repository manager such as \emph{Artifactory} could make the process easier.

\paragraph{GitHub Actions}
GitHub Actions, the CI/CD pipeline for GitHub, was another subject brought up.
Participants mentioned there has been an uptick in vulnerabilities in GitHub Actions.
In response to that, another participant mentioned GitHub's static analysis tool, CodeQL\footnote{https://codeql.github.com/}, now has native support for scanning GitHub Action files.
Tools have also been developed to analyze GitHub Action vulnerabilities, such as zizmor\footnote{https://github.com/woodruffw/zizmor}.

\section{Culture}
\label{sec:culture}

Attendees discussed how they address and foster a culture of security within their organizations.
While security is often evaluated in relation to compliance standards, participants noted that motivating security practices through compliance is challenging.
%Attendees discussed educating developers and leaders in the organization to foster a security mindset.

\subsection{Continued Themes}

\paragraph{Impacts of EO}
Similar to past summits, participants noted that executive orders on software supply chain security are motivating changes within companies.
Executives understand the impact of not meeting the requirements.
One participant noted that as the economy gets bad, security gets overlooked.
They followed this comment up with a suggestion that executive compensation plans should include security as a business outcome.

\subsection{New Ideas}

\paragraph{Fostering a Security Mindset}
Multiple participants discussed that the key to improving security is not to focus on how bad a company is currently doing from a compliance perspective.
Nobody appreciates being criticized. 
Instead, the key is to focus on how the company can improve.
A participant noted that blameless post-mortems are essential to culture.

Many participants shared the \emph{crawl, walk, run} approach, by integrating improvements gradually, for example: picking 1 to 3 things a team can improve on every 3 months.
One participant shared their experience with implementing this approach with workshops.
The workshops help teams decide on what practices they can improve.
Team members can write down 3 to 5 things they want to improve on, and the team can review the entire list from all members to finalize 1 to 3 things to focus on for this quarter.
Scheduling workshops periodically helps teams make small improvements every quarter and fosters a security culture.

\paragraph{Security Across the Entire Organization}
Participants discussed whether company leadership focuses more on business and undermines security.
Some participants shared that the issue is not leadership not recognizing security, but not understanding the time and effort required to implement security practices.
%To address this, a participant recommended involving company leadership in the conversations on improving security, which allows leadership take security into consideration when making business decisions.
%
A participant noted that the problem usually is not the top or the bottom. 
It is everyone else in the middle. 
For example, teams would commit. 
In places where it did not work, it is almost always because the business would not allow them the time to do it. 
Managers began mandating that someone from the business be present in the room when deciding on practices. 
The participant sugguested start these meetings with ``going to commit to'' end them with ``does every one support.'' 
It is important to make sure the business owner is involved in this discussion. 
Adopting this approach caused completion to go from 40\% to 90\%.

A participant asked the question, \textit{Are there operational structures that are more conducive to adopt?}
Another participant noted that at every company, there is at least one team that wants to do it the right way. 
Start there and spread that goodness throughout.
However, they cautioned that one red flag is heavy outsourced product development.
This approach does not work as well in that case.

\section{LLMs and Security}
\label{sec:llm-security}

Recent advancements in AI technology, particularly the emergence of LLMs, have led to widespread adoption by companies and developers alike.
While this technology appears to be very useful, it is still in the early stages of its development lifecycle. 
We asked participants how they leverage the recent advances in AI and whether there are policies around the use of these tools.

Some participants shared the opinion that LLMs are a train that cannot be stopped, noting that we should embrace them and learn how to utilize them.
Others noted that LLMs are becoming commoditized, and given the similarity of APIs, companies are not establishing loyalty ot any specific LLM.
Multiple participants shared that their organizations are still learning how to integrate LLMs and what standards or guidelines should be implemented.
Overall, participants shared the opinion that LLMs are in their infancy.
They can assist developers with tedious tasks and organizations should embrace their use.  However developers should be aware of limitations and concerns with LLMs.

\subsection{Continuing Trends}

\paragraph{Introducing Vulnerabilities}
% devs writing vulnerable code
% are AI generated code evaluated by humans
A participant shared their viewpoint that LLMs have the knowledge of the most experienced engineer, but the maturity of the most junior engineer.
Due to this, some participants are concerned with AI-generated code, including vulnerabilities.
Participants shared that organizations want all AI-generated code to be evaluated by humans to address the risk of introducing vulnerabilities. However, the large amount of code requiring evaluation requires significant manual effort from developers.
% can only used approved LLMs, not a random model on hugginface
Other participants shared that they are only allowed to use organization-approved LLMs.
More popular models are less susceptible to vulnerabilities, whereas newer models from Hugging Face could be more susceptible to vulnerabilities.

\paragraph{Improving Software Testing}
% write unit tests
Unit tests are a testing technique where software components are tested to ensure they function correctly.
Writing unit tests is often a manual process for developers. Developer experience can impact the quality of the unit tests.
Multiple participants shared that LLMs are helpful with writing unit tests.
% help with fuzzing
Similarly, fuzz testing —a software testing technique where various types of input are given to a program to uncover vulnerabilities or program crashes —can also benefit.
Participants shared that LLMs are helpful with coming up with inputs to feed into the program and have increased fuzzing coverage by two or three times.
% detecting obfuscation
Participants also shared that LLMs can be useful for detecting obfuscation in the code.

\subsection{New Ideas}

\paragraph{Software Engineering and Maintenance}
% help upgrade of new versions of libraries
Apart from software testing, LLMs are also helpful with removing technical debt, such as resolving dependencies.
When there is a need to update libraries, LLMs can help determine which version to update to.
% look at contracts from suppliers, identifying deviations, looking at responses from questionnaires sent to suppliers and assigning risk scores
Participants noted that the text summarization ability of LLMs is useful when  looking at contracts from suppliers and providing suggestions for contract renewal.
A participant also mentioned success in using LLMs to aid in threat modeling.
Overall, participants discussed tasks LLMs are helpful with, mostly relating to software engineering and maintenance.

\paragraph{Licenses and Intellectual Property}
% let microsoft or copilot take the risk for licenses
Participants mentioned paying for more popular models and offloading the burden of dealing with vulnerabilities and licenses to the companies maintaining the LLMs.
% being used as training data
However, some participants raised concerns about using a model maintained by another company.
They shared concerns with their data being used as training data to the model and leaking organization information.
Some participants shared that their organizations have their own LLM to address this risk.

\paragraph{Prompt Injection}
One participant observed that ``You can't get in without a badge, but the LLM gets access to everything.''
The participant further noted that ``prompt injection is \emph{your} problem."
They encouraged companies to consider LLMs when evaluating the policy boundaries of a system.
Specifically, APIs traditionally define trust boundaries.
Companies should revisit their API trust assumptions when integrating LLMs into their systems.

\section{Executive Summary}
\label{sec:summary}

%The March 2025 S3C2 Secure Software Supply‑Chain Summit brought together practitioners from 17 organizations to discuss first‑hand experiences with defending modern software supply chains. Six discussion topics; SBOMs, compliance, malicious commits, build infrastructure, culture, and LLMs, revealed a set of unifying themes that underscore the evolving nature of supply chain security challenges. These themes include: (1) the pervasive trust deficit requiring zero trust approaches across the entire development lifecycle; (2) the inherent tension between compliance requirements and practical implementation, particularly with legacy systems; (3) the critical balance between automated security measures and human judgment; (5) the necessity of incremental "crawl, walk , run" approaches to security improvement; and (5) the role of organizational culture and leadership engagement in driving meaningful security practices. Participants across all discussions highlighted that effective supply chain security requires not just technical solutions but also economic incentives for upstream maintainers, cross organizational collaboration, and recognition thaat security is an ongoing process rather than a fixed destination. These insights suggest that the future of software supply chain security lies not in isolated technical fixes but in holistic approaches that address the complex ecosystem in which modern software is developed, distributed, and maintained.

The March 2025 S3C2 Secure Software Supply Chain Summit brought together practitioners from 17 organizations to discuss first‑hand experiences with defending modern software supply chains. 
Six discussion topics; SBOMs, compliance, malicious commits, build infrastructure, culture, and LLMs, revealed a set of unifying themes that underscore the evolving nature of supply chain security challenges. 
All of the panels continued the discussion of themes raised in prior summits, though these discussions often highlighted new nuances and anecdotes.

The panels also revealed the continued maturation of the industry, highlighting new ideas and changes in viewpoint.
For the \emph{SBOM} and \emph{Compliance} panels, there was a sense of frustration towards the many actors in the industry who are only doing enough to meet compliance and no more.
If continued, this may undermine trust of information communicated via SBOMs.
However, individuals in the room were still committed to the vision, and their discussions led to desires for data that provides greater insight into software projects (e.g., contextual SBOMs generated by monitoring the build process).
There was also a growing sense of change to norms.
Companies are beginning to be more open to sharing SBOMs despite initial intellectual property concerns.
The \emph{Malicious Commits} panel revealed that while malicious commits are a significant concern, the difficulty of attributing intent continues to lead to an uncertainty with how best to deal with the threat, and whether or not this is value in making significant changes to existing processes.
The relatively recent attack on the XZ Utils library was still on the minds of participants, as malicious modification of build scripts was incorporated into the discussion.
The \emph{Build Infrastructure} panel revealed perhaps the greatest maturation of the industry since we began running the summits.
The tooling in this space has grown significantly, and the desire to have greater transparency of the build process was not only greater, but also seemed more within reach.
The growing use of AI models also motivated a discussion of how existing build security practices can be adopted to ensure the provenance and integrity of their incorporation into software products.
The \emph{Culture} panel contained the most new discourse of all of the panels.
Companies have been experimenting with different ways of changing company culture, and participants shared successes and strategies with proven track records.
Finally, the \emph{LLMs and Security} panel reflected current trends in the space, highlighting both the potential for LLMs to aid traditional software engineering and maintenance tasks, but also indicating significant caution is being taken over which models are used.
Given the similarity of APIs, LLMs are being commoditized, and companies are not establishing loyalty to any specific LLM.

\section{Acknowledgements}
A big thank you to all Summit participants. We are very grateful for the opportunity to hear about your valuable experiences and suggestions. The Summit was organized by Laurie Williams and William Enck and recorded by Elizabeth Lin and Jonah Ghebremichael.
This material is based upon work supported by the National Science Foundation Grant Nos. 2207008, 2206859, 2206865, and 2206921.
These grants support the Secure Software Supply Chain Summit (S3C2), consisting of researchers at North Carolina State University, Carnegie Mellon University, University of Maryland, and George Washington University. 
Any opinions expressed in this material are those of the author(s) and do not necessarily reflect the views of the National Science Foundation.
%\end{acks}

\bibliographystyle{ACM-Reference-Format}
\bibliography{literature}

\appendix

\section{Initial Discussion Questions}
\label{questions}
\begin{enumerate}

\item \textbf{Software Bill of Materials (SBOM).} Where are you in your journey toward producing an SBOM?  Where are you in your journey toward consuming/using the SBOMs of components and products you use?  What challenges have you faced in SBOM production or use and how have you tried to overcome these challenges?  Are you creating a VEX?  How?

\item \textbf{Compliance.}  What standards do you follow and/or use for guidance for software development practice adoption to reduce software supply chain risk?  (Examples:  SSFD, NIST 800-161, SLSA, S2C2F) How is compliance going (e.g. producing SBOM or self-attestation)?  

\item \textbf{Malicious commits.}  How can malicious commits be detected? What do you think signals a suspicious/malicious commit?  What role does the ecosystem play in detecting malicious commits?

\item \textbf{Build Infrastructure.}  What is being done (or should be being done) to secure the build and deploy process/tooling pipeline (a.k.a SLSA practices)?  Are you working toward reproducible builds?  Do you run your own build server or cloud services?  Do those who use GitHub actions use self-runners?  Are you seeing increased attention by security researchers and attackers on weaknesses in GitHub actions?

\item \textbf{Culture.} What changes have you made to support supply chain security/executive order compliance?  What do you think is needed for nurturing such a security-benefiting  culture?

\item \textbf{LLMs and Supply Chain Security.} How are you leveraging the recent advances in ML/AI in securing your software supply chain?  What are policies around the use of code generation tools in your company?  Do you have a process for choosing safe/secure third-party AI models?

\end{enumerate}

\end{document}